\documentclass[referee]{raa}            

\usepackage{graphicx,times}             
\usepackage{natbib}
\usepackage{amssymb,amsmath}
\usepackage[T1]{fontenc}
\bibpunct{(}{)}{;}{a}{}{,}

\begin{document}

   \title{Photometric and Spectroscopic Study of Two Low Mass Ratio Contact Binary Systems: CRTS J225828.7-121122 and CRTS J030053.5+230139 
}

   \volnopage{Vol.0 (20xx) No.0, 000--000}      
   \setcounter{page}{1}          

   \author{Surjit S. Wadhwa
      \inst{1}
      \and Jelena Petrovi\'c
        \inst{2}
       \and Nick F. H. Tothill
        \inst{1}
   \and Ain Y. De Horta
      \inst{1}
   \and Miroslav D. Filipovi\'c
      \inst{1}
        \and Gojko Djura\v sevi\'c
        \inst{2}
   }

   \institute{School of Science, Western Sydney University, Locked Bag 1797, Penrith, NSW 2751, Australia; {\it 19899347@student.westernsydney.edu.au}\\
        \and
             Astronomical Observatory, Volgina 7, 11060 Belgrade, Serbia\\
\vs\no
   {\small Received~~20xx month day; accepted~~20xx~~month day}}

\abstract{ The study reports photometric and spectroscopic observations of two recently recognised contact binary systems. Both systems show total eclipses and analysis of the light curves indicate both have a very low mass ratios of less than 0.3. We derive absolute parameters from colour and distance based calibrations and show that although both have low mass ratios they are likely to be in a stable orbit and unlikely to merge. In other respects both systems have characteristics similar to other contact binaries with the secondary larger and brighter than main sequence counterparts and we also find that the secondary is considerably denser than the primary in both systems.  
\keywords{binaries: eclipsing -- stars: mass-loss -- techniques: photometric}
}

   \authorrunning{Wadhwa et al}            
   \titlerunning{Low Mass Ratio CBs}  

   \maketitle

%
%
\section{Introduction}           
\label{sect:intro}

As the number of catalogued contact binaries grows with each new sky survey opportunities for their study has also been enhanced. In particular, interest in low mass ratio contact binaries is intense since the recognition that V1309 Sco (= Nova Sco 2008) was in fact a red nova resulting from the merger of contact binary components \citep{2011A&A...528A.114T}. Although a large number of low mass ratio contact binaries have been identified \citep{2021MNRAS.502.2879G, 2022MNRAS.512.1244C, 2022AJ....164..202L, 2023MNRAS.519.5760L} only a handful meet the theoretical criteria for orbital stability \citep{2022RAA....22j5009W, 2023PASP..135g4202W}. \citet{2022JApA...43...94W} recently outlined an efficient method for identifying low mass ratio contact binaries from survey based light curves without formal analysis. We use the methods described in \citet{2022JApA...43...94W} to select two potential low mass ratio contact binaries for follow up ground based observations.

CRTS J225828.7-121122 (C2258) ($\alpha_{2000.0} = 22\ 58\ 28.73$, $\delta_{2000.0} = -12\ 11\ 22.3$) (= ASAS J225829-1211.3,  ASASSN-V J225828.64-121121.9) was recorded as a contact binary system by the All Sky Automated Survey (ASAS) \citep{2002AcA....52..397P} with maximum V band magnitude of 13.37 and period of 0.25208 days. CRTS J030053.5+230139 (C0300) ($\alpha_{2000.0} = 03\ 00\ 53.60$, $\delta_{2000.0} = 23\ 01\ 39.2$) (= ASAS J030054+2301.7, ASASSN-V J030053.58+230138.7) was also identified by ASAS with brightest V magnitude of 13.46 and period of 0.363298 days. Based on the relationships described by \citet{2022JApA...43...94W} we estimated that both systems are likely to be of low mass ratio and as such carried out follow up ground based observations and analysis. In addition, both systems were also observed by the Transiting Exoplanet Survey Satellite (TESS) mission. TESS photometry records the amplitude of 0.50 magnitude for C2258 and 0.42 magnitude for C0300.

\section {Observations}

\subsection{Photometric Observations}
C2258 was observed over 3 days in August 2022 with the Las Cumbres Observatory (LCO) network of 0.4m telescopes using a SBIG STL-6303 CCD camera and Bessel V and B filters. In total 217 images were acquired in V band and an additional 40 images were acquired in B band during total eclipses to document the B-V value. The AstroImageJ \citep{2017AJ....153...77C} package was used to perform differential photometry. TYC 5816-873-1 was the comparison star and 2MASS 22581807-1207490 was the check star. Comparison and check star magnitudes were taken from the Photometric All-Sky Survey \citep{2015AAS...22533616H} calibrations. We find the brightest magnitude to be slightly fainter than survey data at 13.47 magnitude with no significant difference in the two maxima. The primary eclipse has a magnitude of 13.99 indicating an amplitude of 0.52 magnitudes while the secondary eclipse is slightly brighter at 13.94 magnitude. Both eclipses are total of approximately 36 minutes duration. The mean B band magnitude during the primary and secondary eclipses were 14.72 and 14.66 respectively yielding a B-V value of 0.72 for the system. Based on the single observed minima and available survey based V band photometry we revise the orbital elements as follows:

\begin{center}
    
    $HJD_{min} = 2459821.71100(\pm0.00021) + 0.2520799(\pm0.000050)E$\\                   

\end{center}

\noindent C0300 was similarly observed with the LCO network of 0.4 telescopes over 4 nights in November 2022. In total 268 images were acquired in V band and an additional 44 images were acquired in the B band to document the B-V value. As per C2258 differential photometry was performed with the AstroImage J package using 2MASS 03003069 +2259527 as the comparison star and 2MASS 03004569 +2256105 as the check star. Similar to C2258, both maxima are of similar magnitude at 13.51. Both eclipses are total with the primary eclipse slightly fainter at 14.02 magnitude indicating an amplitude of 0.51 magnitude and the secondary eclipse marginally brighter at 14.00 magnitude. Eclipse duration is similar to C2258 at approximately 32 minutes. The B band magnitudes at primary eclipse (14.8) and secondary eclipse (14.78) yield a B-V value 0.78 magnitude. Based on our single observed minima and available survey based V band photometry we revise the orbital elements as follows:
\begin{center}
    
    $HJD_{min} = 2459912.370060(\pm0.0.000577) + 0.36329667(\pm0.000005)E$\\                   

\end{center}
We note that the observed amplitude of C2258 is very similar to the TESS amplitude differing by 0.02 magnitude however the observed amplitude for C0300 is almost 0.1 magnitude greater compared to the TESS photometry. The discrepancy is likely due to blending of the TESS images from nearby fainter stars. TESS images are collected using four small aperture wide-field cameras with very wide point spread function (PSF) of approximately 1 arcmin \citep{2015ApJ...809...77S} significantly increasing the chance of potential blending. A much more dramatic example of blending induced reduction in amplitude of a totally eclipsing contact binary leading to an erroneous light curve solution was described by \citet{2023AN....34420066W} where a system was reported to be of extreme low mass ratio based on SuperWASP \citep{2006PASP..118.1407P} photometry, which also has a resolution of approximately 1 arcmin, was in fact a stable high mass ratio  system with significantly higher amplitude. The difference in amplitude is reflected in the light curve solutions (see below) however we believe the ground based observations in this case yield the more accurate results.

Assessment of period variation, especially when small, requires high cadence long term (over many decades) observations. Given the lack of suitable historical observations (most survey data has a cadence of many days between single observations) no meaningful Observed-Computed $(O-C)$ analysis could be performed for either system. Instead we use the technique employing periodic orthogonal polynomials and an analysis of variance statistic (a quality of fit marker) to fit multiple overlapping subsets, each of approximately 100 - 150 observations, of $V, g'$ and $R$ band data from various surveys including ASAS,  All Sky Automated Survey - Super Nova (ASAS-SN) \citep{2014ApJ...788...48S, 2020MNRAS.491...13J}, Catalina survey \citep{2017MNRAS.469.3688D} and Zwicky Transient Facility (ZTF) \citep{2019PASP..131a8003M} for each system to estimate any significant period variations. The SuperWASP \citep{2006PASP..118.1407P} data for C0300 proved unsuitable due to significant scatter and high reported errors. This is not surprising given brightness of C0300 is near the faint magnitude limit for the survey. The methodology is described in detail by \citet{1996ApJ...460L.107S} and was used by \citet{2011A&A...528A.114T} to demonstrate the exponential decay in the period of V1309 Sco, the only confirmed contact binary merger system. For C2258 there is a slight linear decrease in period of $-1.58\times10^{-7}$ days per year while for C0300 there is a linear increase in period of $3.03\times10^{-7}$ days per year. The period trends are illustrated in Figure 1.

\begin{figure}
    \label{fig:Fig1}
	\includegraphics[width=\columnwidth]{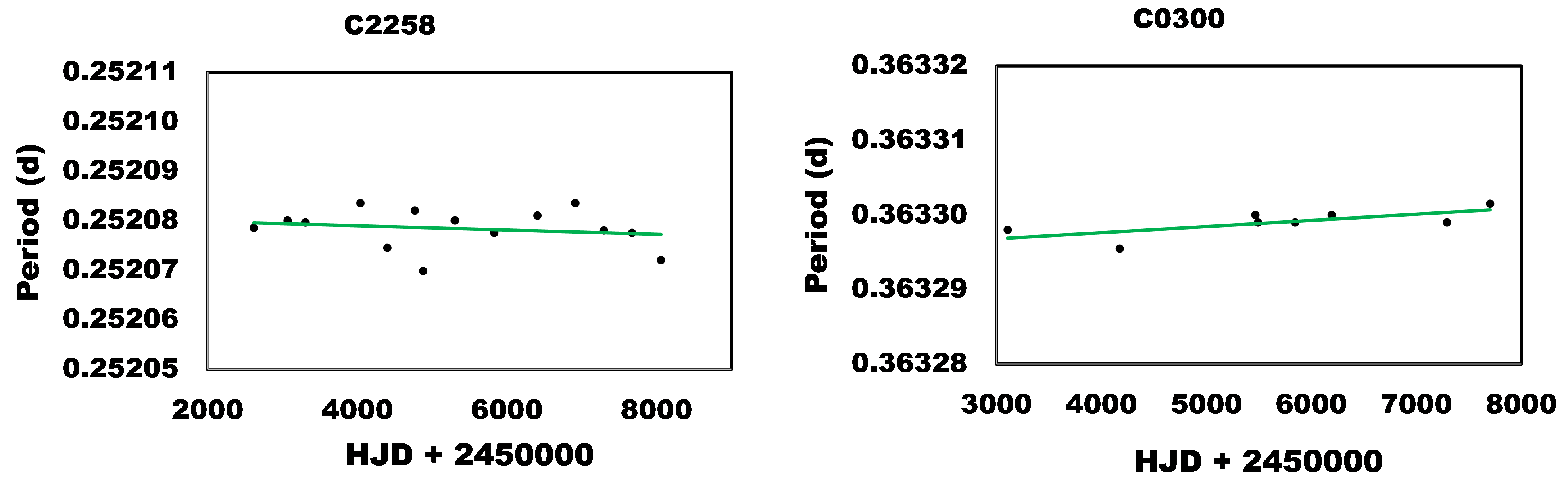}
    \caption{Period variation trend for C2258 (left) and C0300 (right). The green line represents the line of best fit.}
    \end{figure}

\subsection{Effective Temperature and Spectroscopic Observations}

Analysis of contact binary light curves can be successfully carried out where the light curve demonstrates total eclipses \citep{2005Ap&SS.296..221T}. During analysis, the primary component temperature ($T_1$) is usually fixed and non adjustable. The shape of contact binary light curves is almost completely dependent on the Roche geometry and parameters such as the inclination ($i$), mass ratio ($q$), and the Roche equipotential (= degree of contact -- $f$) \citep{1993PASP..105.1433R, 2001AJ....122.1007R, 2022JApA...43...94W}. Although the light curve shape places a tight constrain on the component temperature ($T_2/T_1$) ratio the absolute value of $T_1$ has little influence on the determined geometric elements \citep{1993PASP..105.1433R, 2001AJ....122.1007R}. Previously colour based estimates for $T_1$ have been used however these have proven to be troublesome given the wide variation in catalogued values based on different colours and calibrations. In the case of C2258 the ViziR catalogue service has a range of 5388K to 5923K for the effective temperature of the system and for C0300 the range is even wider at 5076K to 6001K. Although no standard calibration has been adopted many investigators are moving to low resolution spectral observation and spectral class calibration as a means of assigning the effective temperature \citep[see e.g.][]{2022MNRAS.517.1928G, 2022PASJ...74.1421C, 2018PASJ...70...87Z}. We acquired a low resolution spectrum of C2258 using the FLOYD spectrograph attached to the 2 meter telescopes of the LCO network. The spectrum of C2258 was visually matched to library spectra of main sequence stars \citep{1998PASP..110..863P, 1984ApJS...56..257J} to determine it's spectral class as G7.  C0300 was observed by The Large Sky Area Multi-Object Fiber Spectroscopic Telescope (LAMOST) with a reported spectral class of G1 \citep{2018yCat.5153....0L}. The acquired and library spectrum of C2258 and the LAMOST and library spectrum of C0300 are shown in the left panel of Figure 2. We used the April 2022 update of \citet{2013ApJS..208....9P} calibration tables for main sequence stars to determine the temperature of the primary components based on the determined spectral class. The assigned temperature for C2258 was 5550K and for C0300 5860K. The values are within the wide catalogued ranges for both systems.

To further confirm the utility of spectral classification based estimation of the effective temperature we compared the values against a collective photometric approach which collates the photometric data from various pass-bands to construct a single Spectral Energy Distribution (SED) \citep{2008A&A...492..277B}. The SED is then fitted against synthetic theoretical spectra using ${\chi}^2$ minimisation along with Kurucz atmospheric models \citep{2008A&A...492..277B} to determine the effective temperature. The SED estimated effective temperature for C2258 was $5500K$ and $5750K$ for C0300. Given the good agreement between photometric and spectral classification in estimating the effective temperature we consider the use of spectral classification as a valid approach to assign the temperature of the primary component and we have followed this in this study. The SED and fitted model are shown in the right panel of Figure 2 

\begin{figure}
    \label{fig:Fig2}
	\includegraphics[width=\columnwidth]{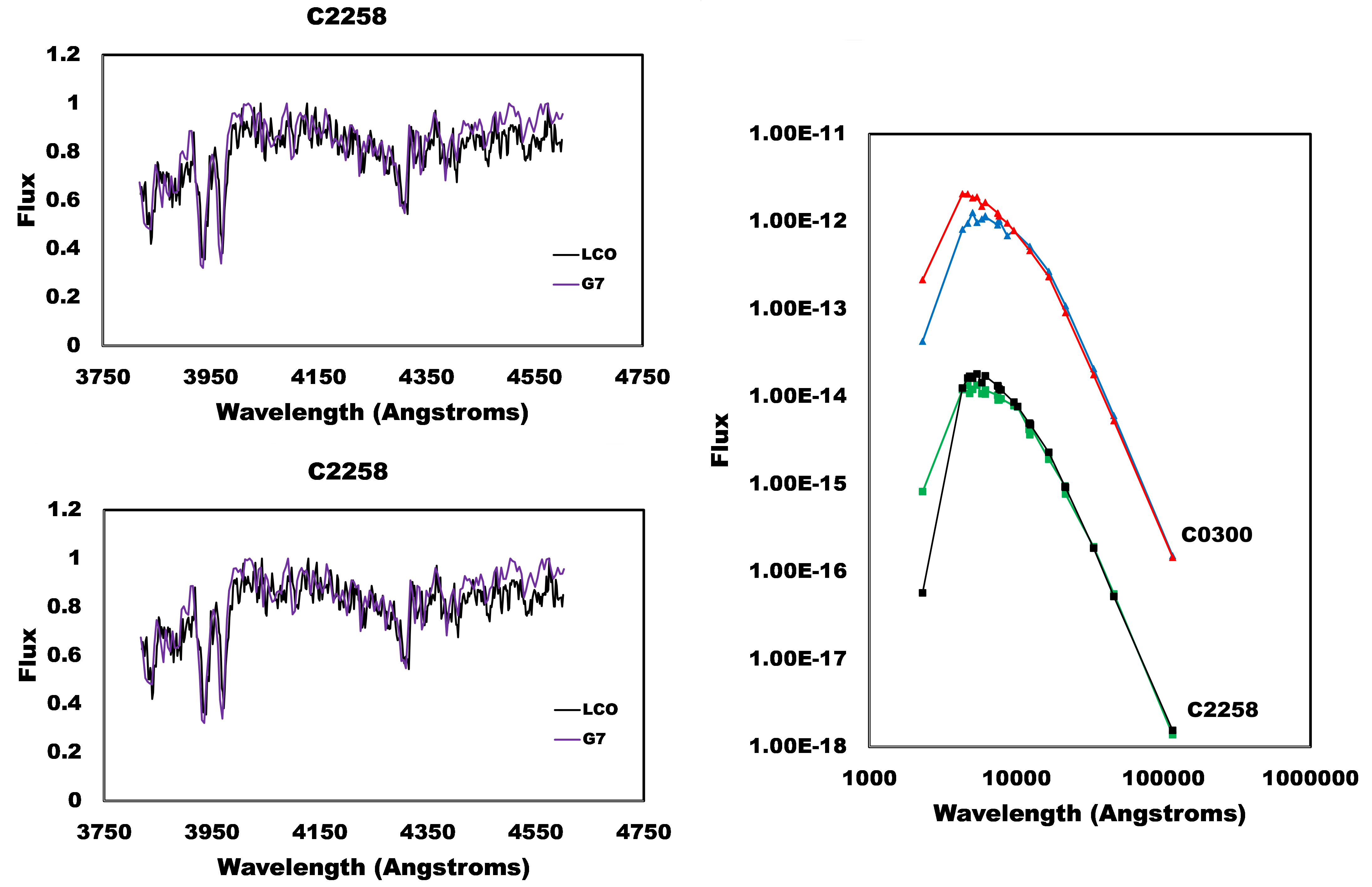}
    \caption{Left Panel: LCO and LAMOST with standard library spectra fro C2258 and C0300\\
    Right Panel: Observed (green and blue) and modelled (black and red) SEDs for C0300 and C2258. The vertical flux is in $erg/cm^2/s/$\AA  and has been shifted vertically by $\times 10^2$ for C0300 for clarity.}
    \end{figure}


\section{Light Curve Solutions and Physical Parameters}

As noted above both systems exhibit total eclipses and thus likely to yield accurate light curve solutions. We used the 2013 version of the Wilson-Devenney code to analyse the V band photometry data. We employ the tried and tested q search/grid method to obtain the mass ratio and complete light curve solution for each system. As there is no appreciable difference in the two maxima for either system only unspotted solutions were modelled. As is usual for low temperature systems the bolometric albedos and gravity darkening coefficients were fixed at ($A_1=A_2 = 0.5$) and ($g_1 = g_2 = 0.32$) respectively and simple reflection treatment applied. We interpolated limb darkening coefficients from \citet{1993AJ....106.2096V} (2019 update). The filter used by the TESS mission is quite broad ranging from 600nm to 1000nm centered near the Ic band (786.5nm). We used the limb darkening co-efficients centered on the Ic band for analysing the TESS photometry. Fitted and observed light curves and three dimensional representations of the contact binaries are shown in Figure 3. The light curve solutions is summarised in Table 1.

\begin{figure}
    \label{fig:fig3}
	\includegraphics[width=\columnwidth]{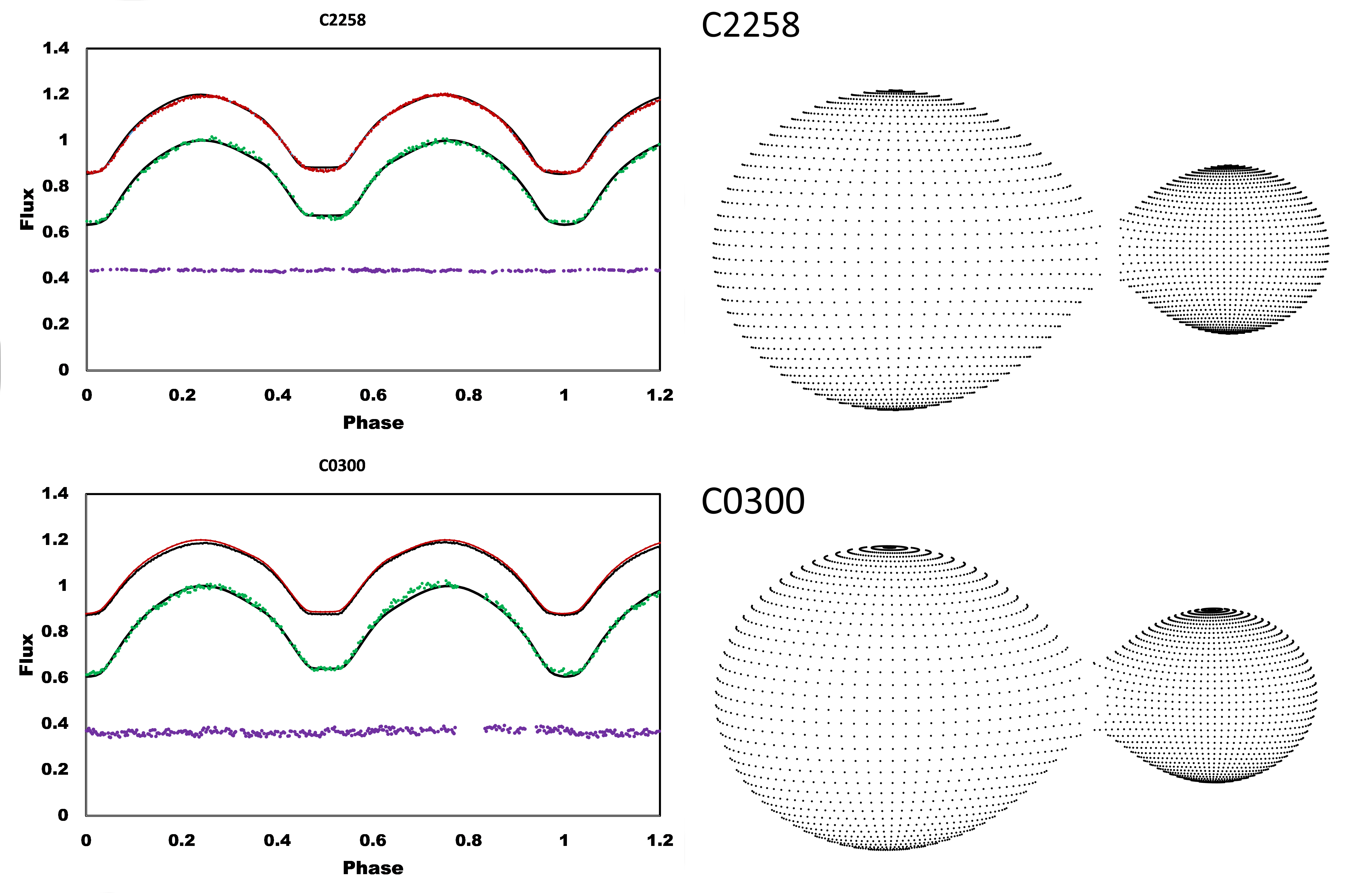}
    \caption{Observed (Green), TESS (Red), fitted (Black) and check (Purple) light curves for C2258 and C0300 along with corresponding 3D representations. The TESS curves have been shifted vertically for clarity.}
    \end{figure}

The light curve solution provides the mass ratio, fractional radii of the components among other geometric parameters. To fully determine the absolute physical parameters one needs to estimate the mass of primary ($M_1$). There is no direct method however secondary calibrations have proven effective. For this study we adopt the mean of an infrared colour ($J-H$) calibration and an absolute magnitude calibration based on distance. We obtained the ($J-H$) colour for both systems from the The Two Micron All Sky Survey (2MASS) \citep{2006AJ....131.1163S} and we used the calibration tables of \citet{2013ApJS..208....9P} ( April 2022 update) for low mass ($0.6M_{\sun} < M < 1.4M_{\sun}$) stars to interpolate the mass of the primary component. 

As the mass ratios are well below 1 the apparent magnitude of the primary component is equal to the apparent magnitude at secondary eclipse. We can use this along with the distance to estimate the absolute magnitude of the primary component. The apparent magnitude of the primary component was first corrected for reddening and distance as follows: Reddening at infinity $E(B-V)_{\infty}$ was determined from \citet{2011ApJ...737..103S} dust maps. This value was then distance scaled $E(B-V)_d$ based on the GAIA DR3 \citep{2022A&A...658A..91A} distance for each system as per the equation \citep{2008MNRAS.384.1178B}:

\begin{equation}
    E(B-V)_d = E(B-V)_{\infty}\Bigg[1-exp\bigg(-\frac{|dsinb|}{h}\bigg)\bigg]
\end{equation}

In the equation $d$ is the GAIA distance in parsecs, $b$ is the galactic latitude, and $h$ is the galactic scale height, adopted as $h=125pc$ \citep{2008MNRAS.384.1178B}.  The $E(B-V)_d$ for C2258 and C0300 were determined as 0.031 and 0.183 respectively and the distance corrected extinctions as 0.1 and 0.57 magnitude respectively. The absolute magnitude of the systems was then determined using the standard distance modulus. We again used the updated \citet{2013ApJS..208....9P} tables for low mass stars to interpolate the mass of the primary based on absolute magnitude. We adopted the mean of the infrared and absolute magnitude calibration for the mass of the primary for subsequent calculations. As the distance based determinations yielded the highest potential errors these were adopted and propagated.

Having determined $M_1$ (and $M_2$ based on the mass ratio $q$) we can use Kepler's third law to estimate the current separation ($A$) between the components. Fractional radii of the components are provided in three orientations by the light curve solution, the geometric mean of these ($r_{1,2}$) can be employed to determine the absolute component radii ($R_{1,2}$) as follows: $R_{1,2} = A(r_{1,2}$). 

Utilising the mean fractional radii we can express the difference in component densities as follows:

\begin{equation}
    \Delta\rho = \frac {0.0189q}{r_2^3(1+q)P^2} - \frac {0.0189}{r_1^3(1+q)P^2}
\end{equation}

Some researchers argue that contact binary evolution may actually result in a change in component designation such that the current primary may have been the initial secondary while the current secondary the initial primary that has lost mass to the current primary. The end result of such a scenario is that the current secondary is rich in core like or heavier elements and the current primary rich in lighter elements. Researchers have 
argued that the densities of the components therefore must differ such that the secondary will always be denser than the primary and $\Delta\rho$ will always be less than zero which is confirmed in both cases \citep{2004A&A...414..317K}. The physical parameters are summarised in Table 1.

A different approach to estimating absolute parameters is to use standard black body emissions based on the determined luminosity as described in \citep{2021AJ....162...13L}. Using this approach with our determined values of the absolute magnitude of the primary we derive $M_1$, $M_2$, $R_1$ and $A$ as $0.74M_{\sun}$, $0.18M_{\sun}$, $0.71R_{\sun}$ and $1.63R_{\sun}$ for C2258 and $0.98M_{\sun}$, $0.28M_{\sun}$, $0.97R_{\sun}$ and $2.32R_{\sun}$ for C0300. We prefer to use estimates based on the geometric light curve solution principally due to the dependency of the black body estimates on the estimated effective temperature. As noted above the catalogued effective temperatures of the two systems described vary by many hundreds of degrees. A change in the effective temperature of $\pm200K$ for C2258 leads to a range in $M_1$ from $0.67M_{\sun}$ to $0.8M_{\sun}$ and for C0300 from $0.94M_{\sun}$ to $1.03M_{\sun}$. As noted above the absolute value of the effective temperature has no significant influence on the geometric light curve solution such that small variations in temperature will not significantly effect absolute parameter estimates determined using light curve solution elements. In addition to the dependency on the effective temperature, the black body based estimate requires multiple steps such as determination of luminosity then radius and then mass of the component. Each step requires error propagation potentially reducing the confidence in the estimated value. Lastly, methodology based on the geometric solution incorporates the distorted morphology of contact binary systems by utilising the geometric mean of the fractional radii in different orientations as opposed a spherical configuration for black body estimates. As noted by \citep{2022JApA...43...94W} that although the primary components of contact binary systems in general follow main sequence characteristics their radii are usually somewhat larger as is reflected in this report. 

\begin{table}
\footnotesize
    \centering
           \begin{tabular}{|l|l|l|l|l|}
    \hline
         Parameter & C2258 & C2258 (TESS)& C0300& C0300 (TESS) \\ \hline
         $T_1$ (K) (Fixed) & 5550&5500&5860 &5860  \\
         $T_2$ (K) & $5536\pm15$&$5540\pm10$ & $5846\pm16$&$5956\pm10$  \\
        Inclination ($^\circ$) & $88.3\pm1.7$&$87.4\pm0.6$ &$83.2\pm0.7$&$80.4\pm0.1$ \\
        Fill-out (\%) & $20\pm3$&$23\pm3$ &$28\pm4$&$15\pm4$  \\
        $r_1$ (mean) &0.516 &0.518 &0.505&0.517 \\
        $r_2$ (mean) &0.273&0.275&0.292&0.269 \\
        $q$ ($M_2/M_1$) & $0.240\pm0.005$ &$0.240\pm0.010$& $0.287\pm0.003$&$0.233\pm0.002$ \\
        $A$/$R_{\sun}$ & $1.71\pm0.01$& & $2.28\pm0.02$& \\
        $M_1/M_{\sun}$ &$0.85\pm0.02$&  & $0.94\pm0.02$&  \\
        $M_2/M_{\sun}$ &$0.21\pm0.01$ & & $0.27\pm0.01$& \\
        $R_1/R_{\sun}$ &$0.88\pm0.02$& &$1.15\pm0.02$& \\
        $R_2/R_{\sun}$ &$0.47\pm0.01$& &$0.67\pm0.02$&  \\
        $M_{V1}$ &$5.92\pm0.08$& &$4.83\pm0.06$&\\
        $\Delta\rho$ &-1.20&&-0.43&\\
        \hline
    \end{tabular}
    \caption{Light curve solution and physical parameters of C2258 and C0300}
\end{table}

At present there is significant interest in orbital stability of contact binary systems. \citet{2021MNRAS.501..229W} recently defined simplified relationships between light curve geometric elements and the mass of the primary with orbital instability. They showed that for low mass primaries the instability mass ratio ($q_{inst}$) is between:

\begin{equation}
\label{eq:qinst-f0}
  q_{inst}=0.0772M_{1}^2-0.3003M_{1}+0.3237 (f=0).  
\end{equation}
and
\begin{equation}
\label{eq:qinst-f1}
    q_{inst}=0.1269M_{1}^2-0.4496M_{1}+0.4403 (f=1)
\end{equation} 

The equation represent the extremes of the instability mass ratio at marginal contact ($f=0$) and complete over contact ($f=1$).

Based on the equations above the instability mass ratio range for C2258 is 0.124 - 0.15 and for C0300 0.11 - 0.13. The light curve solution indicates mass ratios well above the instability mass ratio range suggesting that both systems are likely stable and not merger candidates.

\section{Summary and Conclusion}
Contact binaries present as ideal systems for the study of not only stellar evolutionary scenarios but also orbital dynamics. Orbital stability has received considerable attention recently \citep{2021MNRAS.501..229W, 2022MNRAS.512.1244C, 2023MNRAS.519.5760L} and it is clear from earlier works that contact binaries are likely to become unstable at low mass ratios \citep{2007MNRAS.377.1635A, 2009MNRAS.394..501A}. Based on some simplified criteria \citep{2022JApA...43...94W} we selected two systems showing total eclipses on survey photometry (hence suitable for light curve analysis) that also were likely to have a low mass ratio. Our results confirm that both systems are indeed of low mass ratio at less than 0.3 however once we estimate the physical parameters such as the mass of the primary it is clear that both systems are not near the instability parameters and as such not potential merger candidates. The study does however confirm that recently published selection and analysis relationships can be easily implemented to selectively observe low mass ratio contact binary systems.

\section*{Acknowledgements}
 
This research has made use of the VizieR catalogue access tool, CDS, Strasbourg, France (DOI : 10.26093/cds/vizier).\\

\noindent This research has made use of the SIMBAD database, operated at CDS, Strasbourg, France.\\

\noindent This work makes use of observations from the Las Cumbres Observatory global telescope network.\\

\noindent This publication makes use of VOSA, developed under the Spanish Virtual Observatory (https://svo.cab.inta-csic.es) project funded by MCIN/AEI/10.13039/501100011033/ through grant PID2020-112949GB-I00. VOSA has been partially updated by using funding from the European Union's Horizon 2020 Research and Innovation Programme, under Grant Agreement number 776403 (EXOPLANETS-A).\\

\noindent J. P. and G. D.  gratefully  acknowledge financial support of the Ministry of Education, Science and Technological Development of the Republic of Serbia through contract No. 451-03-9/2021-14/200002.

\bibliographystyle{raa}
\bibliography{bibtex}

\begin{thebibliography}{40}
\providecommand\natexlab[1]{#1}
\providecommand\JournalTitle[1]{#1}

\bibitem[{Anders} {et~al.}(2022)]{2022A&A...658A..91A}
{Anders}, F., {Khalatyan}, A., {Queiroz}, A.~B.~A., {et~al.} 2022, \aap, 658,
  A91

\bibitem[{Arbutina}(2007)]{2007MNRAS.377.1635A}
{Arbutina}, B. 2007, \mnras, 377, 1635

\bibitem[{Arbutina}(2009)]{2009MNRAS.394..501A}
{Arbutina}, B. 2009, \mnras, 394, 501

\bibitem[{Bayo} {et~al.}(2008)]{2008A&A...492..277B}
{Bayo}, A., {Rodrigo}, C., {Barrado Y Navascu{\'e}s}, D., {et~al.} 2008, \aap,
  492, 277

\bibitem[{Bilir} {et~al.}(2008)]{2008MNRAS.384.1178B}
{Bilir}, S., {Ak}, S., {Karaali}, S., {et~al.} 2008, \mnras, 384, 1178

\bibitem[{Chang} {et~al.}(2022)]{2022PASJ...74.1421C}
{Chang}, L.-F., {Zhu}, L.-Y., {Sarotsakulchai}, T., \& {Soonthornthum}, B.
  2022, \pasj, 74, 1421

\bibitem[{Christopoulou} {et~al.}(2022)]{2022MNRAS.512.1244C}
{Christopoulou}, P.-E., {Lalounta}, E., {Papageorgiou}, A., {et~al.} 2022,
  \mnras, 512, 1244

\bibitem[{Collins} {et~al.}(2017)]{2017AJ....153...77C}
{Collins}, K.~A., {Kielkopf}, J.~F., {Stassun}, K.~G., \& {Hessman}, F.~V.
  2017, \aj, 153, 77

\bibitem[{Drake} {et~al.}(2017)]{2017MNRAS.469.3688D}
{Drake}, A.~J., {Djorgovski}, S.~G., {Catelan}, M., {et~al.} 2017, \mnras, 469,
  3688

\bibitem[{Gazeas} {et~al.}(2021)]{2021MNRAS.502.2879G}
{Gazeas}, K.~D., {Loukaidou}, G.~A., {Niarchos}, P.~G., {et~al.} 2021, \mnras,
  502, 2879

\bibitem[{Guo} {et~al.}(2022)]{2022MNRAS.517.1928G}
{Guo}, D.-F., {Li}, K., {Liu}, F., {et~al.} 2022, \mnras, 517, 1928

\bibitem[{Henden} {et~al.}(2015)]{2015AAS...22533616H}
{Henden}, A.~A., {Levine}, S., {Terrell}, D., \& {Welch}, D.~L. 2015, in
  American Astronomical Society Meeting Abstracts, Vol. 225, American
  Astronomical Society Meeting Abstracts \#225, 336.16

\bibitem[{Jacoby} {et~al.}(1984)]{1984ApJS...56..257J}
{Jacoby}, G.~H., {Hunter}, D.~A., \& {Christian}, C.~A. 1984, \apjs, 56, 257

\bibitem[{Jayasinghe} {et~al.}(2020)]{2020MNRAS.491...13J}
{Jayasinghe}, T., {Stanek}, K.~Z., {Kochanek}, C.~S., {et~al.} 2020, \mnras,
  491, 13

\bibitem[{K{\"a}hler}(2004)]{2004A&A...414..317K}
{K{\"a}hler}, H. 2004, \aap, 414, 317

\bibitem[{Li} {et~al.}(2022)]{2022AJ....164..202L}
{Li}, K., {Gao}, X., {Liu}, X.-Y., {et~al.} 2022, \aj, 164, 202

\bibitem[{Li} {et~al.}(2021)]{2021AJ....162...13L}
{Li}, K., {Xia}, Q.-Q., {Kim}, C.-H., {et~al.} 2021, \aj, 162, 13

\bibitem[{Liu} {et~al.}(2023)]{2023MNRAS.519.5760L}
{Liu}, X.-Y., {Li}, K., {Michel}, R., {et~al.} 2023, \mnras, 519, 5760

\bibitem[{Luo} {et~al.}(2018)]{2018yCat.5153....0L}
{Luo}, A.~L., {Zhao}, Y.~H., {Zhao}, G., \& {et al.} 2018, VizieR Online Data
  Catalog, V/153

\bibitem[{Masci} {et~al.}(2019)]{2019PASP..131a8003M}
{Masci}, F.~J., {Laher}, R.~R., {Rusholme}, B., {et~al.} 2019, \pasp, 131,
  018003

\bibitem[{Pecaut} \& {Mamajek}(2013)]{2013ApJS..208....9P}
{Pecaut}, M.~J., \& {Mamajek}, E.~E. 2013, \apjs, 208, 9

\bibitem[{Pickles}(1998)]{1998PASP..110..863P}
{Pickles}, A.~J. 1998, \pasp, 110, 863

\bibitem[{Pojmanski}(2002)]{2002AcA....52..397P}
{Pojmanski}, G. 2002, \actaa, 52, 397

\bibitem[{Pollacco} {et~al.}(2006)]{2006PASP..118.1407P}
{Pollacco}, D.~L., {Skillen}, I., {Collier Cameron}, A., {et~al.} 2006, \pasp,
  118, 1407

\bibitem[{Rucinski}(1993)]{1993PASP..105.1433R}
{Rucinski}, S.~M. 1993, \pasp, 105, 1433

\bibitem[{Rucinski}(2001)]{2001AJ....122.1007R}
{Rucinski}, S.~M. 2001, \aj, 122, 1007

\bibitem[{Schlafly} \& {Finkbeiner}(2011)]{2011ApJ...737..103S}
{Schlafly}, E.~F., \& {Finkbeiner}, D.~P. 2011, \apj, 737, 103

\bibitem[{Schwarzenberg-Czerny}(1996)]{1996ApJ...460L.107S}
{Schwarzenberg-Czerny}, A. 1996, \apjl, 460, L107

\bibitem[{Shappee} {et~al.}(2014)]{2014ApJ...788...48S}
{Shappee}, B.~J., {Prieto}, J.~L., {Grupe}, D., {et~al.} 2014, \apj, 788, 48

\bibitem[{Skrutskie} {et~al.}(2006)]{2006AJ....131.1163S}
{Skrutskie}, M.~F., {Cutri}, R.~M., {Stiening}, R., {et~al.} 2006, \aj, 131,
  1163

\bibitem[{Sullivan} {et~al.}(2015)]{2015ApJ...809...77S}
{Sullivan}, P.~W., {Winn}, J.~N., {Berta-Thompson}, Z.~K., {et~al.} 2015, \apj,
  809, 77

\bibitem[{Terrell} \& {Wilson}(2005)]{2005Ap&SS.296..221T}
{Terrell}, D., \& {Wilson}, R.~E. 2005, \apss, 296, 221

\bibitem[{Tylenda} {et~al.}(2011)]{2011A&A...528A.114T}
{Tylenda}, R., {Hajduk}, M., {Kami{\'n}ski}, T., {et~al.} 2011, \aap, 528, A114

\bibitem[{van Hamme}(1993)]{1993AJ....106.2096V}
{van Hamme}, W. 1993, \aj, 106, 2096

\bibitem[{Wadhwa} {et~al.}(2023{\natexlab{a}})]{2023PASP..135g4202W}
{Wadhwa}, S.~S., {Arbutina}, B., {Tothill}, N. F.~H., {et~al.}
  2023{\natexlab{a}}, \pasp, 135, 074202

\bibitem[{Wadhwa} {et~al.}(2021)]{2021MNRAS.501..229W}
{Wadhwa}, S.~S., {De Horta}, A., {Filipovi{\'c}}, M.~D., {et~al.} 2021, \mnras,
  501, 229

\bibitem[{Wadhwa} {et~al.}(2022{\natexlab{a}})]{2022RAA....22j5009W}
{Wadhwa}, S.~S., {De Horta}, A., {Filipovi{\'c}}, M.~D., {et~al.}
  2022{\natexlab{a}}, Research in Astronomy and Astrophysics, 22, 105009

\bibitem[{Wadhwa} {et~al.}(2022{\natexlab{b}})]{2022JApA...43...94W}
{Wadhwa}, S.~S., {De Horta}, A.~Y., {Filipovi{\'c}}, M.~D., {et~al.}
  2022{\natexlab{b}}, Journal of Astrophysics and Astronomy, 43, 94

\bibitem[{Wadhwa} {et~al.}(2023{\natexlab{b}})]{2023AN....34420066W}
{Wadhwa}, S.~S., {DeHorta}, A.~Y., {Filipovi{\'c}}, M., \& {Tothill}, N. F.~H.
  2023{\natexlab{b}}, Astronomische Nachrichten, 344, e20220066

\bibitem[{Zhou} {et~al.}(2018)]{2018PASJ...70...87Z}
{Zhou}, X., {Qian}, S., {Boonrucksar}, S., {et~al.} 2018, \pasj, 70, 87

\end{thebibliography}


\begin{thebibliography}{37}
\providecommand\natexlab[1]{#1}
\providecommand\JournalTitle[1]{#1}

\bibitem[{Boyle} {et~al.}(2000)]{Boyle+etal+2000}
{Boyle}, B.~J., {Shanks}, T., {Croom}, S.~M., {et~al.} 2000, \mnras, 317, 1014

\bibitem[{Casali} {et~al.}(2007)]{Casali+etal+2007}
{Casali}, M., {Adamson}, A., {Alves de Oliveira}, C., {et~al.} 2007, \aap, 467,
  777

\bibitem[{Chiu} {et~al.}(2007)]{Chiu+etal+2007}
{Chiu}, K., {Richards}, G.~T., {Hewett}, P.~C., \& {Maddox}, N. 2007, \mnras,
  375, 1180

\bibitem[{Constantin} {et~al.}(2002)]{Constantin+etal+2002}
{Constantin}, A., {Shields}, J.~C., {Hamann}, F., {Foltz}, C.~B., \& {Chaffee},
  F.~H. 2002, \apj, 565, 50

\bibitem[{Croom} {et~al.}(2004)]{Croom+etal+2004}
{Croom}, S.~M., {Smith}, R.~J., {Boyle}, B.~J., {et~al.} 2004, \mnras, 349,
  1397

\bibitem[{Fan} {et~al.}(2001{\natexlab{a}})]{Fan+etal+2001b}
{Fan}, X., {Narayanan}, V.~K., {Lupton}, R.~H., {et~al.} 2001{\natexlab{a}},
  \aj, 122, 2833

\bibitem[{Fan} {et~al.}(2001{\natexlab{b}})]{Fan+etal+2001a}
{Fan}, X., {Strauss}, M.~A., {Schneider}, D.~P., {et~al.} 2001{\natexlab{b}},
  \aj, 121, 54

\bibitem[{Hambly} {et~al.}(2008)]{Hambly+etal+2008}
{Hambly}, N.~C., {Collins}, R.~S., {Cross}, N.~J.~G., {et~al.} 2008, \mnras,
  384, 637

\bibitem[{Hewett} {et~al.}(2006)]{Hewett+etal+2006}
{Hewett}, P.~C., {Warren}, S.~J., {Leggett}, S.~K., \& {Hodgkin}, S.~T. 2006,
  \mnras, 367, 454

\bibitem[{Hu} {et~al.}(2008)]{Hu+etal+2008}
{Hu}, C., {Wang}, J.-M., {Ho}, L.~C., {et~al.} 2008, \apj, 687, 78

\bibitem[{Kong} {et~al.}(2006)]{Kong+etal+2006}
{Kong}, M.-Z., {Wu}, X.-B., {Wang}, R., \& {Han}, J.-L. 2006, \cjaa, 6, 396

\bibitem[{Lawrence} {et~al.}(2007)]{Lawrence+etal+2007}
{Lawrence}, A., {Warren}, S.~J., {Almaini}, O., {et~al.} 2007, \mnras, 379,
  1599

\bibitem[{Maddox} {et~al.}(2008)]{Maddox+etal+2008}
{Maddox}, N., {Hewett}, P.~C., {Warren}, S.~J., \& {Croom}, S.~M. 2008, \mnras,
  386, 1605

\bibitem[{Markwardt}(2009)]{Markwardt+2009}
{Markwardt}, C.~B. 2009, in Astronomical Society of the Pacific Conference
  Series, Vol. 411, Astronomical Data Analysis Software and Systems XVIII, ed.
  D.~A. {Bohlender}, D.~{Durand}, \& P.~{Dowler}, 251

\bibitem[{Richards} {et~al.}(2002)]{Richards+etal+2002}
{Richards}, G.~T., {Fan}, X., {Newberg}, H.~J., {et~al.} 2002, \aj, 123, 2945

\bibitem[{Richards} {et~al.}(2009)]{Richards+etal+2009}
{Richards}, G.~T., {Myers}, A.~D., {Gray}, A.~G., {et~al.} 2009, \apjs, 180, 67

\bibitem[{Schlegel} {et~al.}(1998)]{Schlegel+etal+1998}
{Schlegel}, D.~J., {Finkbeiner}, D.~P., \& {Davis}, M. 1998, \apj, 500, 525

\bibitem[{Schmidt}(1963)]{Schmidt+1963}
{Schmidt}, M. 1963, \nat, 197, 1040

\bibitem[{Schneider} {et~al.}(2010)]{Schneider+etal+2010}
{Schneider}, D.~P., {Richards}, G.~T., {Hall}, P.~B., {et~al.} 2010, \aj, 139,
  2360

\bibitem[{Shen} {et~al.}(2011)]{Shen+etal+2011}
{Shen}, Y., {Richards}, G.~T., {Strauss}, M.~A., {et~al.} 2011, \apjs, 194, 45

\bibitem[{Smith} {et~al.}(1994)]{Smith+etal+1994}
{Smith}, J.~D., {Djorgovski}, S., {Thompson}, D., {et~al.} 1994, \aj, 108, 1147

\bibitem[{Smith} {et~al.}(2005)]{Smith+etal+2005}
{Smith}, R.~J., {Croom}, S.~M., {Boyle}, B.~J., {et~al.} 2005, \mnras, 359, 57

\bibitem[{Su} {et~al.}(1998)]{Su+etal+1998}
{Su}, D.~Q., {Cui}, X., {Wang}, Y., \& {Yao}, Z. 1998, in Society of
  Photo-Optical Instrumentation Engineers (SPIE) Conference Series, Vol. 3352,
  Society of Photo-Optical Instrumentation Engineers (SPIE) Conference Series,
  ed. L.~M. {Stepp}, 76

\bibitem[{Tsuzuki} {et~al.}(2006)]{Tsuzuki+etal+2006}
{Tsuzuki}, Y., {Kawara}, K., {Yoshii}, Y., {et~al.} 2006, \apj, 650, 57

\bibitem[{Vestergaard} \& {Peterson}(2006)]{Vestergaard+Peterson+2006}
{Vestergaard}, M., \& {Peterson}, B.~M. 2006, \apj, 641, 689

\bibitem[{Vestergaard} \& {Wilkes}(2001)]{Vestergaard+Wilkes+2001}
{Vestergaard}, M., \& {Wilkes}, B.~J. 2001, \apjs, 134, 1

\bibitem[{Warren} {et~al.}(2000)]{Warren+etal+2000}
{Warren}, S.~J., {Hewett}, P.~C., \& {Foltz}, C.~B. 2000, \mnras, 312, 827

\bibitem[{Wright} {et~al.}(2010)]{Wright+etal+2010}
{Wright}, E.~L., {Eisenhardt}, P.~R.~M., {Mainzer}, A.~K., {et~al.} 2010, \aj,
  140, 1868

\bibitem[{Wu} \& {for the LAMOST Extragalactic Survey Team}(2011)]{Wu+2011}
{Wu}, X.-B., \& {for the LAMOST Extragalactic Survey Team}. 2011,
  arXiv:1111.0738

\bibitem[{Wu} {et~al.}(2012)]{Wu+etal+2012}
{Wu}, X.-B., {Hao}, G., {Jia}, Z., {Zhang}, Y., \& {Peng}, N. 2012, \aj, 144,
  49

\bibitem[{Wu} \& {Jia}(2010)]{Wu+Jia+2010}
{Wu}, X.-B., \& {Jia}, Z. 2010, \mnras, 406, 1583

\bibitem[{Wu} {et~al.}(2011)]{Wu+etal+2011}
{Wu}, X.-B., {Wang}, R., {Schmidt}, K.~B., {et~al.} 2011, \aj, 142, 78

\bibitem[{Wu} {et~al.}(2004)]{Wu+etal+2004}
{Wu}, X.-B., {Zhang}, W., \& {Zhou}, X. 2004, \cjaa, 4, 17

\bibitem[{Wu} {et~al.}(2010{\natexlab{a}})]{Wu+etal+2010a}
{Wu}, X.-B., {Chen}, Z.-Y., {Jia}, Z.-D., {et~al.} 2010{\natexlab{a}}, Research
  in Astronomy and Astrophysics, 10, 737

\bibitem[{Wu} {et~al.}(2010{\natexlab{b}})]{Wu+etal+2010b}
{Wu}, X.-B., {Jia}, Z.-D., {Chen}, Z.-Y., {et~al.} 2010{\natexlab{b}}, Research
  in Astronomy and Astrophysics, 10, 745

\bibitem[{York} {et~al.}(2000)]{York+etal+2000}
{York}, D.~G., {Adelman}, J., {Anderson}, Jr., J.~E., {et~al.} 2000, \aj, 120,
  1579

\bibitem[{Zhao} {et~al.}(2012)]{Zhao+etal+2012}
{Zhao}, G., {Zhao}, Y.-H., {Chu}, Y.-Q., {Jing}, Y.-P., \& {Deng}, L.-C. 2012,
  Research in Astronomy and Astrophysics, 12, 723

\end{thebibliography}

\end{document}